# Learn to segment single cells with deep distance estimator and deep cell detector

Weikang Wang[a], David A.Taft[a], Yi-Jiun Chen[a], Jingyu Zhang[a], Callen T. Wallace[b], Min Xu[c], Simon C. Watkins[b], Jianhua Xing[a,d],*

[a]Department of Computational and System Biology, University of Pittsburgh, Pittsburgh, PA, 15260, USA

[b]Department of Cell Biology, and Center for Biologic Imaging, University of Pittsburgh, Pittsburgh, PA 15261 USA

[c]Department of Computational Biology, Carnegie Mellon University, Pittsburgh, PA, 15213, USA

[d]UPMC-Hillman Cancer Center, University of Pittsburgh, Pittsburgh, PA, 15232, USA

Weikang Wang: weikang@pitt.edu
David A.Taft: dat81@pitt.edu
Yi-Jiun Chen: yic42@pitt.edu
Jingyu Zhang: jennifer.in3501@pitt.edu
Callen T. Wallace: callenwall@gmail.com
Min Xu: mxu1@andrew.cmu.edu
Simon C. Watkins: simon.watkins@pitt.edu
Jianhua Xing: xing1@pitt.edu

**ABSTRACT**

Single cell segmentation is a critical and challenging step in cell imaging analysis. Traditional processing methods require time and labor to manually fine-tune parameters and lack parameter transferability between different situations. Recently, deep convolutional neural networks (CNN) treat segmentation as a pixel-wise classification problem and have become a general and efficient method for image segmentation. However, cell imaging data often possesses characteristics that adversely affect segmentation accuracy: absence of established training datasets, few pixels on cell boundaries, and ubiquitous blurry features. We developed a strategy that combines strengths of CNN and traditional watershed algorithm. First, we trained a CNN to learn Euclidean distance transform (EDT) of the mask corresponding to the input images (deep distance estimator). Next, we trained a faster R-CNN (Region with CNN) to detect individual cells in the EDT image (deep cell detector). Then, the watershed algorithm performed the final segmentation using the outputs of previous two steps. Tests on a library of fluorescence, phase contrast and differential interference contrast (DIC) images showed that both the combined method and various forms of the pixel-wise classification algorithm achieved similar pixel-wise accuracy. However, the combined method achieved significantly higher cell count accuracy than the pixel-wise classification algorithm did, with the latter performing poorly when separating connected cells, especially those connected by blurry boundaries. This difference is most obvious when applied to noisy images of densely packed cells. Furthermore, both deep distance estimator and deep cell detector converge fast and are easy to train.

**Keywords**

Convolutional neural networks, watershed, connected cells, blurry boundary, cell count accuracy

# 1. Introduction

Automated segmentation has become an area of focused research for both clinical and basic science applications due to the time and labor these automated efforts could save clinicians and investigators over traditional, manual methods (Arbelle et al., 2018; Drozdzal et al., 2018). Single live cell imaging is a basic method for studying the temporal and spatial dynamics of individual cells. It provides information about cell heterogeneity that is concealed in bulk measurement studies (Mullassery et al., 2008; Muzzey and van Oudenaarden, 2009). Single live cell imaging typically generates massive data, and subsequent analysis and mining from such massive imaging data can be challenging (Meijering and van Cappellen, 2006). Single cell segmentation, which identifies and outlines regions of interest in an image, is one of the most difficult tasks in biomedical image analysis (Su et al., 2013; Uchida, 2013). It is a key step for extracting multiple types of quantitative information, such as fluorescent protein expression, the total number of particles in single cells, and quantitative measurements of cell shape (Kherlopian et al., 2008; Roeder et al., 2012; Uchida, 2013). Several commonly used tools exist for such quantification, such as ImageJ (https://imagej.nih.gov/ij/) and CellProfiler (Carpenter et al., 2006). However, to address a specific segmentation problem, one needs to use a combination of different techniques and manually fine-tune multiple parameters. The process is often time-consuming and labor-intensive. Furthermore, existence of noise or slight variation of the images may lead to a poor segmentation outcome and require additional rounds of tedious parameter tuning (Meijering et al., 2009; Roeder et al., 2012; Uchida, 2013). Consequently, single cell segmentation often suffers from low efficiency with poor reproducibility and transferability.

Machine learning methods, especially deep-learning, have developed quickly in recent years (LeCun et al., 2015). Notably, deep convolutional neural network (CNN) provides a powerful and general method for image classification, segmentation and object detection (Kraus et al., 2016; Krizhevsky et al., 2012; Long et al., 2015; Ren et al., 2015). CNN is good at learning information of compositional hierarchical features from images. The algorithm is trained to learn both intensity and shape features (LeCun et al., 2015). For instance, CNN can be used for object detection by predicting the bounding boxes of objects as well as their categories based on the intensity and shape information (Redmon et al., 2016; Ren et al., 2015).

CNN has also been applied to single cell segmentation (Akram et al., 2016; Ronneberger et al., 2015; Van Valen et al., 2016). In these studies, a basic strategy is to classify individual pixels into intracellular, boundary, and background categories, and train CNN to learn this pixel-wise classification. Compared with traditional segmentation methods, CNN-based pixel-wise classification shows high accuracy and efficiency (Ronneberger et al., 2015; Van Valen et al., 2016). Technical difficulties, however, still exist when segmenting cell images with deep-learning algorithms, especially when cells are densely packed and interact with neighboring cells tightly, a situation common in vivo and in vitro conditions. When segmenting these cells, several challenges affect the accuracy of applying the deep-learning approaches. First, currently there is no established training dataset of cell micrographs that includes multiple types of images (Hilsenbeck et al., 2017; Ronneberger et al., 2015). Therefore, one usually needs to first prepare training data for their specific cell type and imaging condition. Pixel-wise annotation of a large number of cell images for training a segmentation model is difficult and time-consuming. This often restricts one to train CNN with only a small amount of training labels from their experiment data, which may significantly limit the performance of trained CNN models (Van Valen et al., 2016). Second, some cells are physically connected. To segment the connected cells precisely, a CNN needs to identify the boundaries of individual cells precisely. A cell boundary is thin, i.e., it contains much less pixels than the interior of the cell does. Since in existing CNN-based algorithms segmentation is transformed into a pixel-wise classification problem, less pixels mean less labeled training samples. Thus, the problem of few boundary pixels further worsens the above problem of small training data. Third, it is almost unavoidable that in microscopic images some cells or boundaries are blurred due to small depth of field, focus drift, or other limitation of the sample or microscope. Neural networks have also been used for restoration tasks to clean up fluorescent images prior to image analysis in order to solve imaging limitations (Weigert et al., 2018). Accounting for blurry features is key since they may mislead a CNN to erroneously segment multiple connected objects as one object. Such under-segmentation can seriously affect subsequent analyses steps such as intensity calculation and tracking.

Alternatively, watershed is a widely used traditional segmentation method that does not require prior knowledge of images under study or training labels. Notably, watershed performs well when identifying blurry boundaries since it identifies the peak of gray-level change as an edge even if the gray level changes slightly (Uchida, 2013). Thus, watershed can overcome the above three limitations of CNN. Unfortunately, watershed has its own weakness. It is not suitable for processing noisy images. Existence of noise can induce over-segmentation or irregular boundaries segmented with watershed (Roerdink and Meijster, 2000). A commonly used technique to remedy the over-segmentation problem is using markers as the start of flooding in watershed. These markers have different labels and are separated into different objects after segmentation (Roerdink and Meijster, 2000). It is challenging to generate the markers efficiently and accurately.

Therefore, both CNN and the traditional watershed methods have advantages and disadvantages for segmenting densely packed cells. Given that CNN is able to learn the image intensity composition rules that represent objects (LeCun et al., 2015), we proposed that one can use CNN to simplify the original images then process the simplified images with the watershed algorithm to combine the strengths of the two methods. Based on the above intuition, in this work we developed a three-step procedure for single cell segmentation. Instead of training CNN for pixel-wise classification as in existing studies, the new method trains two CNNs to learn alternative cell features and has three notable novelties. First, one trains a CNN (deep distance estimator) to learn the Euclidean distance transform (EDT) instead of pixel classification of the original input images. It converts the original noisy images into simplified ones that can be processed with watershed directly. Second, one trains a faster R-CNN (deep cell detector) to detect individual cell from the EDT images and generate markers for watershed (Ren et al., 2015). The deep cell detector helps to avoid the problem of under-segmentation of pixel-wise classification algorithm and over-segmentation of directly using watershed on original images. Third, one uses watershed to perform final segmentation on the predicted EDT images with the markers generated with deep cell detector. With the inputs, the watershed step is straightforward and does not require exhaustive parameter fine-tuning. Compared with CNN-based direct pixel-wise classification method, our combined method shows similar accuracy in pixel level, but significantly increased cell count accuracy. Expanded and better-curated training data can further improve both pixel accuracy and cell count accuracy of the method.

## 2. METHODS

### 2.1. Cell culture and image acquisition

We cultured three types of cells for imaging. Human T47D cells with endogenous membrane protein E-cadherin fused EGFP (Chen et al., 2017) were cultured in DMEM (Gibco, 11995) with 10% FBS (Gibco, 10437028). Mouse NMuMg cells were cultured in DMEM with 10 µg/ml insulin (Sigma, I0516) and 10% FBS. Human HK2 cells were cultured in DMEM/F12 (Gibco, 11330) medium with 5 µg/ml insulin, 0.02 µg/ml dexamethasone (Sigma, D4902), 0.01 µg/ml selenium (Sigma, S5261), 5 µg/ml transferrin (T8158), and 10% FBS. Cells were seeded at ~30% confluence in MatTek glass bottom culture dish (35 mm) and cultured for 1 day before imaging. We used three representative types of cell images as our test systems, fluorescence images of T47D-E-cadherin-EGFP cells (40× oil objective, N.A.=1.3), differential interference contrast (DIC) images of HK2 cells (20× objective, N.A.=0.75), and phase contrast images of NMuMg cells (20× objective, N.A.=0.45). HK2 cells are also stained with Calcein AM (Invitrogen C1430, FITC) and taken images together with DIC (20× objective, N.A.=0.75). All images were taken with Nikon Ti-E microscope (Andor Neo SCC-00211).

## 2.2. Direct pixel-wise classification method

As CNN is powerful at extracting hierarchies of different features, its usage on image recognition and classification has been developed quickly in recent years (LeCun et al., 2015; LeCun et al., 1989). Long and co-workers proposed fully convolutional networks (FCN) for semantic segmentation, and developed CNN into a general method for image segmentation (Long et al., 2015). Subsequently, several architectures of CNN have been raised for image segmentation (Badrinarayanan et al., 2017; Chen et al., 2014; Noh et al., 2015; Yu and Koltun, 2015). Specifically, CNN and related approaches have been applied for segmenting bio-medical images (Ciresan et al., 2012; Ronneberger et al., 2015; Van Valen et al., 2016). Ronneberger and coworkers proposed U-net for processing phase contrast and DIC images on the basis of fully convolutional networks (Long et al., 2015; Ronneberger et al., 2015). Van Valen *et. al.* proposed DeepCell for segmenting bacteria and mammalian cells from phase contrast images with the assistance of fluorescence images of cell nuclei (Van Valen et al., 2016).

Existing CNN-based segmentation methods transform the segmentation problem into a pixel-wise classification problem (Garcia-Garcia et al., 2017; Long et al., 2015). That is, one divides pixels within an image into different categories, and CNN is trained to learn category classification. For single cell segmentation, pixels are classed into three categories: background (labeled with an integer index 0), intra-cellular pixels (with index 1), and pixels on boundaries (with index 2) (Fig. S1). Trained by a set of pre-categorized images as the ground truth, a CNN reads the original cell images and predicts the mask integer values for individual pixels with the largest probability.

We followed the method of FCN with an encoder and decoder network architecture (Fig. 1) (Badrinarayanan et al., 2017). The encoder network is the same as the convolutional layers of VGG16 except the fully connected layers (Simonyan and Zisserman, 2014). The decoder part contains a hierarchy of decoders that correspond to the encoder layers. To recover the size of the images, we used up-sampling layers in the decoder part.

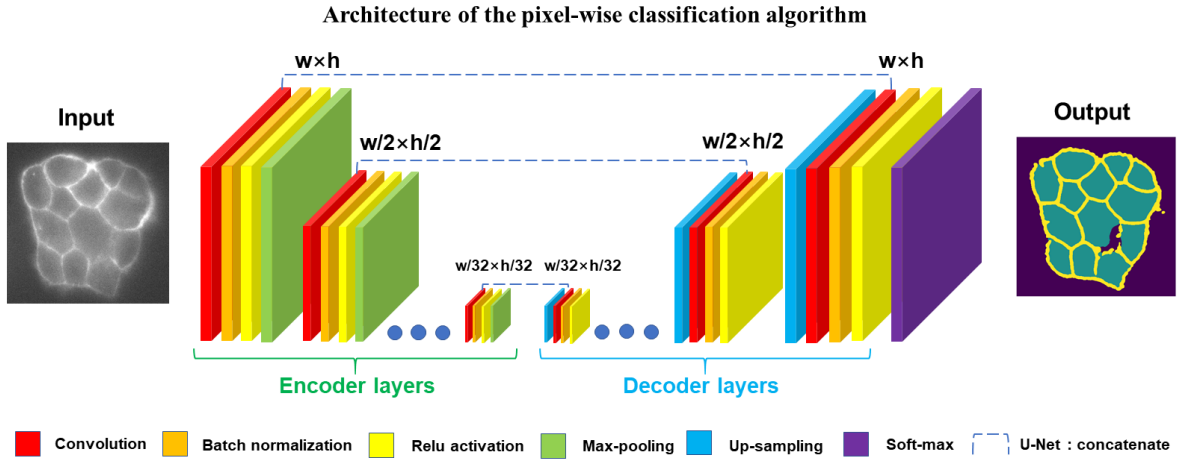

**Figure 1. Architecture of the pixel-wise classification FCN.** The network was trained to classify pixels into one of three categories: background, intra-cellular, or cell boundary.

We also tried concatenating the corresponding encoder and decoder layers following the algorithm in U-net (Fig. 1) (Ronneberger et al., 2015). In the pixel-wise classification FCN, the decoder part is followed by a soft-max classification layer to make pixel-wise prediction (Badrinarayanan et al., 2017). For this task, we used a cross entropy function as the loss function. To deal with unbalanced data, we increased the weight of pixels on cell boundary by using a class weighted cross entropy (CWCE) loss function $loss = -\sum_{i}^{n} \left( P_{i,y_{true}} \times w_i \times \log\left(P_{i,y_{pred}}\right) \right)$. In the formula $n$ is the total number of classes, $P_{i,y_{true}}$ is the true probability of current pixel on $i_{th}$ class, $w_i$ is the weight value of $i_{th}$ class, and $P_{i,y_{pred}}$ is the predicted probability of current pixel on $i_{th}$ class.

We used Adaptive Moment Estimation (*adam*) as the optimizer for the pixel-wise classification FCN (Kingma and Ba, 2014), with a learning rate 0.001, an exponential decay rate for estimation of the first moment $\beta_1$ 0.9, an exponential decay rate for estimation of the second moment $\beta_2$ 0.999, a small number $\varepsilon = 10^{-8}$ to prevent any division by zero, and the learning rate decay 0.

For the training data of the pixel-wise classification FCN, we selected a number of images and manually segmented these images to obtain three-category label mask. We randomly cropped multiple regions with a size of 256 × 256 or 320 × 320 pixels in each of the input images and the corresponding label mask (Fig. S2). We used these cropped patches as the training data. The number of training patches for T47D fluorescence, NMuMg phase contrast and HK2 DIC images are 131, 95, and 357, respectively. Each input training patch was normalized by dividing the median pixel value of its own (Van Valen et al., 2016). While preparing the ground-truth data, we segmented only cell bodies of HK2 cells. Each HK2 cell has its cell body and a thin pseudopodium. The latter is challenging to recognize in DIC images even with human eyes. The CNN is trained to segment cell body only instead of the whole cell (cell body plus pseudopodium). The cell body is profiled based on thresholding and manual correction on Calcein AM fluorescent images. To reduce the connected objects, we also thickened the width of cell boundaries by several pixels, two pixels for T47D fluorescence images and NMuMg phase contrast images and four for HK2 DIC images, to increase the number of pixels on cell boundaries. For convenience of discussions, we refer them as thick data set, and the original unmodified ones as thin data set.

## 2.3. Combined CNN and watershed method with deep distance estimator and deep cell detector

We developed a three-step procedure to combine CNN and watershed (Fig. 2). The basic strategy is to use CNN models to detect individual cells and use watershed to map cell boundaries. In the first step, we transform the classification problem to a regression problem, and train a CNN model (deep distance estimator) to learn the distance transform of the three-category label mask of the segmentation label. Distance transform is a commonly used technique for creating topological surface for the watershed algorithm (Uchida, 2013). In the second step, we train another network to detect individual cells in the images using a faster R-CNN method (Ren et al., 2015). The centers of identified bounding boxes serve as markers in watershed. In the third step, the watershed algorithm takes the outputs from the first step and the second step as input and performs the final segmentation. Below we describe these three steps in detail.

In the first step, we still use the architecture of an encoder-decoder classification FCN as in the direct pixel-wise classification method, except changing the last soft-max layer into a rectified linear unit (RELU) activation layer $f(x) = \max(0, x)$, where x is the input value from the previous layer (Krizhevsky et al., 2012). Other layers are kept as the same. This deep distance estimator is trained to learn the EDT of the three-category label mask corresponding to the input images. The three-category label mask is transformed into a binary mask firstly. In the binary mask, values of pixels in the interior of a cell are set to be one, and those of the exterior pixels (including cell boundary and background) are set to be zero. The EDT of a pixel is defined to be the distance from this pixel to the nearest pixel with a value of zero. The Euclidean distance between two points with Cartesian coordinates $(a_1, b_1)$ and $(a_2, b_2)$ is calculated by the following formula:

$$D = \sqrt{(a_1 - a_2)^2 + (b_1 - b_2)^2}.$$

We use a mean squared error (MSE) loss function:

$$MSE = \frac{1}{n} \sum_{i=1}^{n} \left( y_{pred} - y_{true} \right)^2,$$

where $i$ represents the $i_{th}$ pixel, $n$ is the total number of pixels, $y_{pred}$ is the predicted value of present pixel, and $y_{true}$ is the true value of present pixel.

We again use *adam* as the optimizer of deep distance estimator (Kingma and Ba, 2014). The learning rate was set to be 0.001, $\beta_1$ 0.9, $\beta_2$ 0.999, $\varepsilon$ $10^{-8}$, and the learning rate decay 0.

In the second step, we train a faster R-CNN to detect all the cells in the EDT image obtained from the first step. Faster R-CNN is one of the most popular methods for object detection (Ren et al., 2015). A faster R-CNN contains two parts: a region proposal network (RPN) and a classifier of region of interest (ROI). The RPN gives a prediction on whether a bounding box contains an object or not. The ROI classifier identifies which category an object belongs to and tightens the bounding box generated by RPN. In this work, we only have one object category: cell. The deep cell detector detects cells in the output of the deep distance estimator and predicts the bounding boxes of cells. The loss function and optimizer of faster R-CNN are defined as in the original paper (Ren et al., 2015). A bounding box is defined by its top left corner coordinate ($u_1$, $v_1$) and bottom right corner coordinate ($u_2$, $v_2$). After obtaining all the bounding boxes, we use the centers of bounding boxes as markers in the following watershed segmentation. This step can be skipped if there is cell nucleus staining in the original images. One can easily segment stained cell nuclei with thresholding or other algorithms and use the nuclei as markers in watershed. In some applications, especially in live cell imaging, nucleus staining is not a favorable option due to introduced additional phototoxicity and occupancy of one fluorescence channel.

In the third step, watershed segmentation is performed using the predicted distance transform from the first step and the markers predicted by deep cell detector in the second step. When using the watershed algorithm, one needs to specify the mask of watershed, a binary array of the same shape as the input image. The mask assigns every pixel with a value either "true" or "false". Only pixels with a true value are segmented, and those with a false value are set as background. We applied thresholding on the prediction of distance transform to generate the mask. The values of pixels above the threshold value are set as true and the others are set as false. One can easily fine-tune the threshold value, which is always close to 0. For T47D, HK2 and NMuMg cells, the threshold values used in this paper were set to be 0.2, 0.1 and 0.1, respectively. It should be emphasized that this threshold value is the only parameter that need to be tuned manually for watershed, while it can also be optimized automatically by maximizing the pixel accuracy (see definition below). This is drastically different from a traditional application of watershed, which requires tedious tuning of multiple parameters.

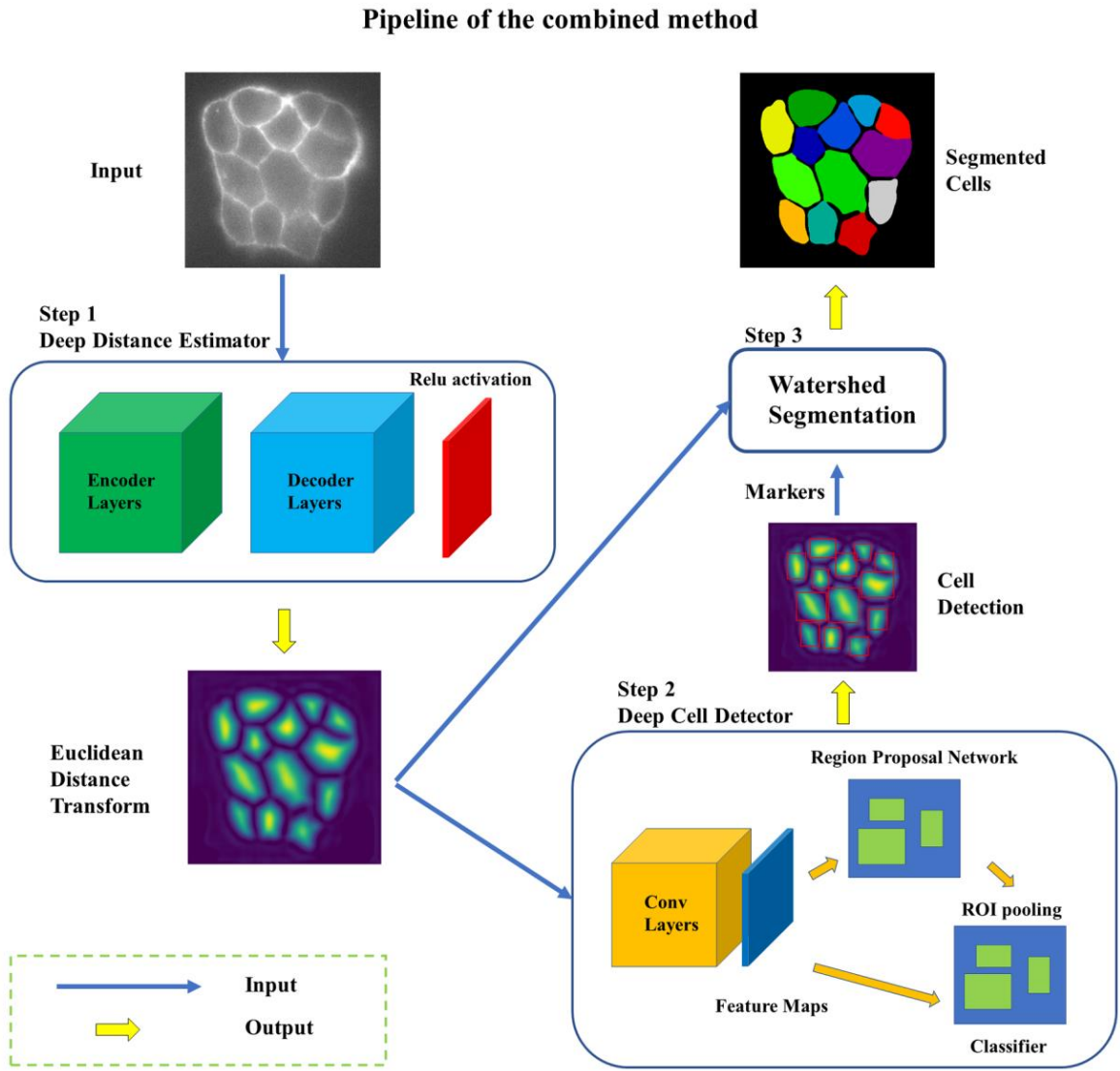

**Figure 2. Pipeline of the proposed method combining CNN and watershed.** The method has three steps, calculating the EDT of the original input images using the deep distance estimator, cell detection on the prediction of EDT with a Deep cell detector (faster R-CNN), and final segmentation with the watershed algorithm on the prediction of EDT with cell detection results as markers.

To generate training data for the deep distance estimator, we used the same cropped patches and generated the corresponding three-category label mask for each patch as in the pixel-wise classification method. The training label of the distance transform is calculated based on the three-category label mask (Fig. S2). Each EDT label patch was normalized by dividing the median pixel value of its own. All the randomly cropped patches are stored into a four-dimensional matrix ($N \times w \times h \times c$, where $N$ is the number of patches, $w$ is the width of a patch, $h$ is the height of a patch and $c$ is the number of channels of a patch. In our case, $c$ is 1). All the corresponding training label patches are also transformed into a four-dimensional matrix ($N \times w \times h \times c$, where $c$ equals 1). The two matrixes served as input and ground truth separately in the deep cell estimator in the training process.

To prepare for the training data of the deep cell detector, we generated bounding boxes based on the distance transform images. We found that detecting whole cells yield low accuracy. So, instead we only detected the core of each cell. First, we identified the pixel with the highest value of distance in each cell. Then we calculated the pixels whose distance to this center pixel are smaller than 2/3 of the major axis length of the cells, with an additional requirement that values of these pixels need to be larger than 1/3 of the value of the center pixel. The bounding boxes were set to enclose the pixels that meet both criteria. The position and size of each bounding box were calculated and written into an xml file corresponding to the cropped patch (following the data structure of PASCAL VOC dataset).

Implementation of our method is on the keras framework (Chollet, 2015). All the codes, training data, testing data and weight files can be found using the following link,

[https://github.com/opnumten/single_cell_segmentation](https://github.com/opnumten/single_cell_segmentation). The code of classification FCN is modified based on the code from [https://github.com/opnumten/keras-segnet](https://github.com/opnumten/keras-segnet). The code of deep cell detector is modified based on the code from [https://github.com/opnumten/keras-frcnn](https://github.com/opnumten/keras-frcnn). Training of all the neural networks was performed on an NVIDIA TITAN X 12GB GPU.

## 2.4. Evaluation of segmentation

We computed the cell count accuracy (CCA) with the following equation (Chalfoun et al., 2014):

$$CCA = \frac{TP}{N + FP}.$$

Where *TP* is the true positive cell count, *N* is the total number of cells in the input images, and *FP* is the false positive cell count. CCA is an important measure of segmentation accuracy, and a low CCA can lead to severe error in subsequent single cell analyses. For comparison, for every type of images we labeled three images each containing ~ 30 - 70 cells and compared the CCA of different algorithms.

The pixel accuracy $\frac{\sum_i n_{ii}}{\sum_i t_i}$ and mean intersection over union (mean IU) $\frac{1}{n_{cl}} \sum_i \frac{n_{ii}}{\left(t_i + \sum_j n_{ji} - n_{ii}\right)}$ were calculated following the algorithm reported in FCN. Here $n_{ij}$ is the number of pixels in class $i$ that are predicted to be class $j$, $t_i = \sum_j n_{ij}$ is the total number of pixels that are predicted to be class $i$, and $n_{cl}$ is the number of classes (Long et al., 2015). We tested the pixel accuracy of different algorithms on three images for every type of image.

## 3. RESULTS

### 3.1. Various types of images show features problematic for direct segmentation.

There are two basic types of live cell imaging, fluorescence-based and non-fluorescence-based (e.g., transmitted light images). Fluorescence labeling provides additional features, such as labeled cell membranes, for aiding cell segmentation. However, using fluorescence imaging for the purpose of cell segmentation increases the effect of phototoxicity and limits the frequency and duration of live cell imaging. It also occupies fluorescence channels of a microscope that can otherwise be used for other purposes. Generating the labeled cells can be labor-intensive and time-consuming. Using transmitted light imaging, such as DIC and phase-contrast, requires no labeling, and can significantly reduce the exposure time and intensity, but the images are expected to be more difficult for segmentation. We tested on both fluorescence images obtained using human T47D cells with EGFP fused to the membrane protein E-cadherin, and two types of transmitted light images, i.e., DIC images using human HK2 cells and phase-contrast images using mouse NMuMG cells (Fig. 3A). T47D and NMuMG cells show typical polygon-shaped epithelial morphology with cells tightly packed together. HK2 cells have less packing density, and neighboring cells are only partially connected. Therefore, using images from these three cell types we can evaluate the effect of cell packing on the accuracy of cell segmentation.

We spotted blurry boundaries in all three types of images, and the red circles in Fig. 3A indicate some of them. A few cells have long segments of their boundaries barely detectable, e.g., a boundary between two cells in region 2. A more common situation is that a few blurry pixels make it difficult to close a boundary. For example, the two cells in region 1 share a boundary that is clear except in close proximity to a "T" intersection. The blurry boundaries impose challenges for segmentation, as we will see below.

Compared to the intra-cellular pixels, the number of pixels on the boundary of a cell is small. For example, in Fig. 3A the thin threads of high intensity fluorescence of E-cadherin EGFP reveal the width of cell boundary. Based on the fluorescence intensity, in Fig. S1A we divided the pixels of the image into those on cell boundary (yellow color) and intra-cellular pixels (green color). The ratio of pixels on cell boundary to intra-cellular pixels is 0.139. As training a CNN requires sufficient amount of training data, the low percentage of pixels on cell boundary may decrease the classification accuracy on this category, especially when the training set is small. One solution for this problem is thickening the cell boundary in the ground truth training data. Thickening the cell boundaries in Fig. S1A by two pixels leads to an increase of the cell boundary to intra-cellular pixel ratio to 0.237 (Fig. S1B). However, there is a compromise on how many pixels to increase, since thickening the boundary with more pixel values affects the accuracy of segmentation. In the following studies, we used both the original thin-boundary data set and augmented thick-boundary data set to train the neural networks.

**3.2. The pixel-wise classification algorithm has low performance on segmenting connecting cells.**

Figure 3B-E show example outputs of pixel-wise classification algorithm applied to different types of images. The CNN trained with thin-boundary training sets (Fig. 3B) failed to recognize most of the connected cells with blurry and incomplete boundaries. Cells within every region circled in Fig. 3A are predicted as a single cell. For example, the program recognized most of boundaries of the two cells in region 1, but still classified them as one cell since it failed to complete the "T" intersection and separate them completely. The CNN trained with thick-boundary training data (Fig. 3C) performed noticeably better when segmenting cells that had only a few blurry boundary pixels, e.g. in regions 3, 5 and 7. However, the program could not accurately segment cells with extended blurry boundaries, e.g., in regions 1, 2, 4 and 6.

Since even the thick-boundary data set has unbalanced numbers of boundary and intracellular pixels, we proposed that using the CWCE loss function may improve the performance of the encoder-decoder classification FCN (Panchapagesan et al., 2016). In this case the class weight values are calculated based on the proportions of pixels in different categories of the training data set. This loss function increases penalty of wrong prediction on boundary pixels, which partially eliminate influence of unbalanced data set. The outputs with the CWCE loss function (Fig. 3D), as well as those of U-net trained with the CWCE loss function (Fig. 3E), showed only slight improvement over those obtained with the thick-boundary data trained CNN. These results suggest that the unbalanced training data set is not a main reason for cell mis-segmentation.

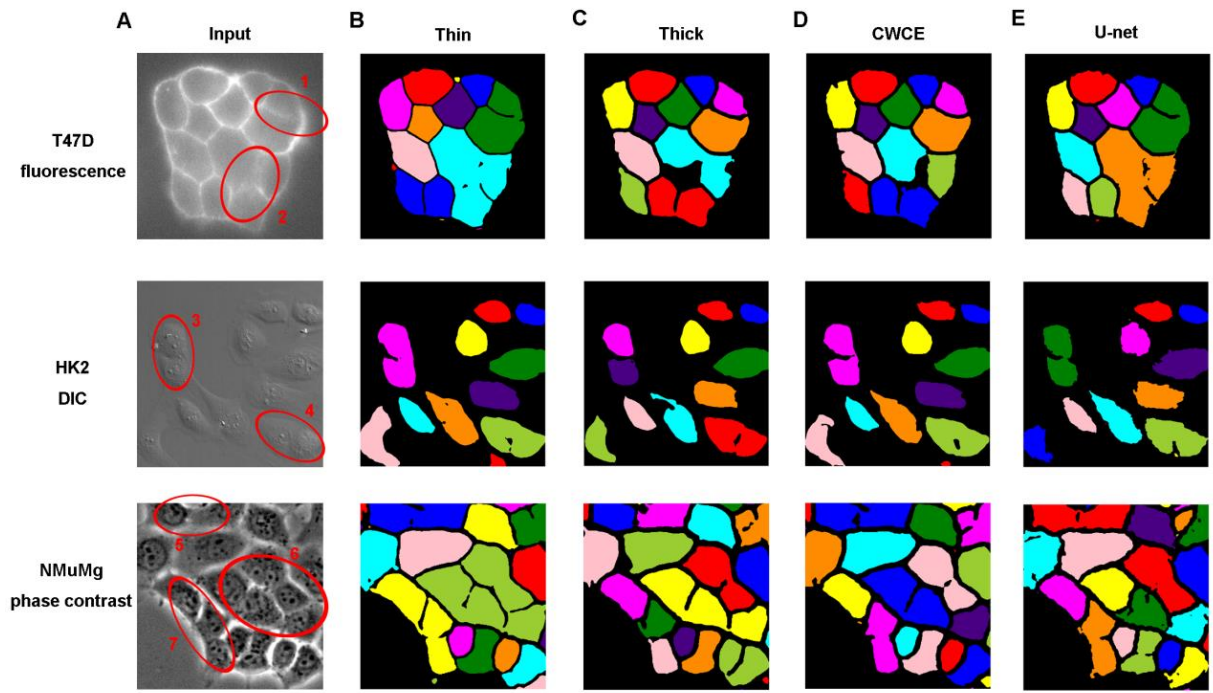

**Figure 3. Example input images and outputs of the pixel-wise classification algorithm.** Red circles highlight cells that have blurry boundaries and are mis-segmented by some or all methods tested here.

Taken together, various forms of the pixel-wise classification algorithm have the tendency of under-segmenting cells with blurry boundaries. The problem is especially severe with densely packed cells, where shared boundaries between neighboring cells are often disrupted by blurry pixels. The connected cells result in low CCA and introduce errors for subsequent single cell analyses, such as cell tracking.

**3.4. Combined CNN and watershed method can accurately segment connected cells**

In our combined CNN and watershed algorithm (Fig. 2), we adopted a different strategy. In step 1 a deep distance estimator is trained to learn EDT. Since the Euclidean distance assumes a continuous instead of a Boolean value, pixels that are close to cell boundary have smaller values. Specifically consider the regions with blurred boundaries in the original images. Since the pixels on the blurred boundaries are on the extension of the clear boundary, values of their EDT are likely to be small. Indeed, in the transform images (Fig. 4B), each cell had a brighter central region representing larger values of the Euclidean distance, which faded while moving towards cell boundaries. In EDT image, some neighboring cells with shared blurry boundaries in the original images, e.g., those in region 3 of Fig. 3A, were well separated. While some were still partially connected in the Euclidean distance representation, such as those in region 1, they did not impose difficulties in step 2, which uses a deep cell detector to detect single cells and predict the corresponding bounding boxes (Fig. 4C). The centers of these bounding boxes serve as markers for watershed and avoid the problem of over-segmentation. Fig. 4D shows that for all three types of images the new method successfully separated the connected cells whose boundaries are blurry.

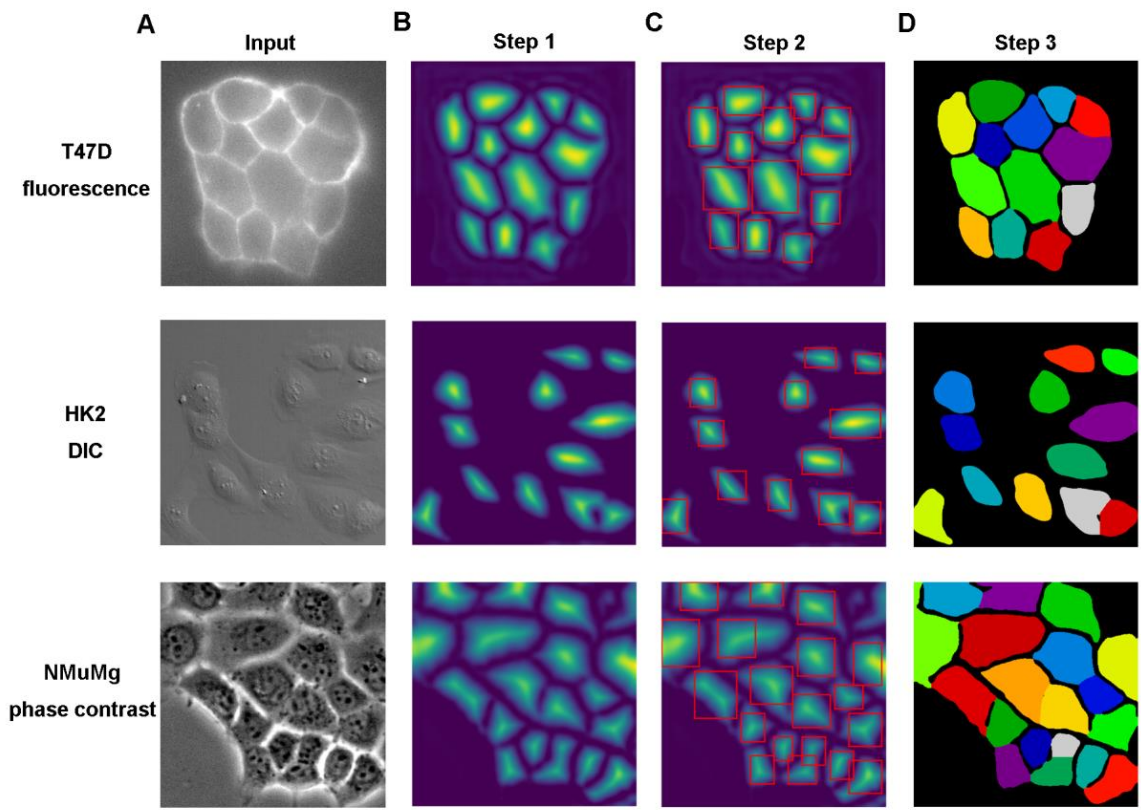

**Figure 4. Example output of the combined method applied to different types of images.** The four columns are input images, predicted EDT, cell detection (bounding box) results with faster R-CNN, and watershed segmentation results, respectively.

## 3.5. The combined method improves CCA significantly over pixel-wise classification methods.

We compared the CCA of various pixel-wise classification FCNs and the combined method (Fig. 5). The CCA of pixel-wise classification FCN trained with thin-boundary data was the lowest in all three types of test images. Furthermore, the algorithm consistently performed better on CCA with sparsely packed (HK2 cells) than those densely packed cells (T47D and NMuMG cells). Sparsely packed cells are mostly not connected to neighboring cells, and thus only have a small fraction of boundaries shared with other cells (Fig. S3A). In contrast, each of densely packed cells has on average 5-6 connected neighbors, and most cells have more than 50% boundaries shared with other cells (Fig. S3B & C). Using thick-boundary training data led to ~10% increase of CCA with sparsely packed cells, and a more dramatic ~ 20% increase with densely packed cells. These results are consistent with what observed in Fig. 3 that thickening boundaries helped to separate connected cells with a few blurry boundary pixels. Neither use of the CWCE loss function nor a combined U-net/CWCE loss function led to further improvement on CCA. Notice that with the thick-boundary training data, all the pixel-wise classification FCNs performed better on fluorescence images that on transmitted light (DIC and phase contrast) images. One possible explanation is that fluorescence images have simpler features and sharper contrast between boundary pixels with fluorescence label and other pixels. In comparison, the combined approach achieved the highest and most consistent CCA in all three types of images. The improvement is especially remarkable with phase-contrast images of densely packed cells with ~80% CCA, where the pixel-wise classification FCNs approach achieved only ~30 – 60% CCA. All the methods showed similar ~ 90% pixel accuracy (Fig. 5B) and ~80% mean IU (Fig. 5C).

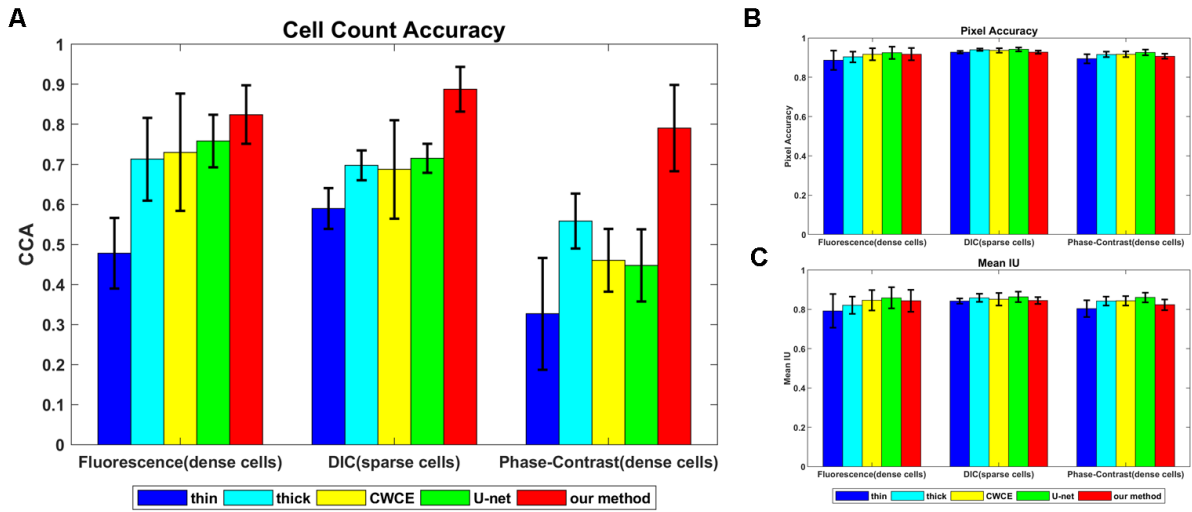

**Figure 5. Comparison between segmentation accuracy achieved by pixel-wise classification FCNs and the combined method on different types of images.** A: CCA; B: pixel accuracy; C: mean IU.

**3.6. Regression filters noises and leads to smooth predicted cell boundaries.**

In our combined method, we transformed the classification problem into a regression problem and trained a FCN to predict the EDT directly from input images in the first step. Alternatively, one can first train a pixel-wise classification FCN to make pixel category prediction with on the input image following the procedure depicted in Fig. 1, then calculate the corresponding EDT (Fig. S4A). For this purpose, one needs to transform the three categories prediction of pixel-wise classification FCN into a binary mask, with values of intra-cellular pixels as 1 and values of boundary and background pixels as 0. Next one performs cell detection and watershed segmentation on the EDT as in our combined method.

Using the encoder-decoder classification FCN trained with thick-boundary data and CWCE loss function, we tested this alternative combined method on phase contrast images (Fig. S4B). Consistent with previous results, the pixel-wise classification output under-segmented connected cells, sometime mis-classified some intracellular pixels as background pixels and incorrectly predicted segmented cells with holes and rugged boundaries (Fig. S4C). An additional step of watershed segmentation indeed improved segmentation of connected cells but showed no improvement in reducing the holes and rugged boundaries. More detrimentally, the method led to over-segmentation of some cells into multiple smaller objects (Fig. S4D red circles). In contrast, our original combined method predicted smooth cell boundaries without much problem of over-segmentation (Fig. S4E). These results suggest that segmentation results can be quite different depending on whether a FCN is trained to learn pixel classifications or EDT. Therefore, we recommend the deep distance estimator in step 1 of the combined method. Furthermore, the convergence speed of training a pixel-wise classification FCN is slower than that of a deep distance estimator (Fig. S5).

## 4. DISCUSSION

**Comparison with pixel-wise classification FCN**

As shown in the results (Fig. 3), it is especially challenging for pixel-wise classification FCN algorithms to segment densely packed cells, which tend to mis-recognize two contacting cells with portions of boundary pixels blurry as one cell. According to our test, the thick-boundary training label improves encoder-decoder FCN's performance on CCA significantly for all three types of images (Badrinarayanan et al., 2017; Long et al., 2015), which is due to the increase of the amount of training labels on minor boundary pixels. Therefore, the way of preparing training labels is critical for the performance of CNN. However, the CWCE loss function, another method to deal with the unbalanced training labels, does not increase CCA (Panchapagesan et al., 2016). The concatenation structure of U-net also fails to improve the CCA (Ronneberger et al., 2015). These pixel-wise classification FCNs are sensitive to noise, and there are even holes inside single cells in the segmentation output (Fig. 3, Fig. S4). In addition, the convergence speed is also slow due to its classification nature (Fig. S5). The combined method developed in this work uses strategies different from the pixel-wise classification algorithms. It shows similar pixel accuracy and mean IU but significant higher CCA (Fig. 5), and it performs well when separating those connected cells with blurry boundaries between them. Furthermore, this method is more tolerant to noise.

**Novelties of the combined method**

The first novelty of the combined method developed here is we trained a CNN to learn the EDT instead of pixel classification of the original input image. The EDT is a quasi-continuous function of the spatial coordinate, and pixel classification is a step function with discrete values and changes abruptly from an intracellular pixel to a neighboring boundary pixel then to a background pixel. Intuitively, one expects that it is less challenging to learn a continuous function than a step function (Fig. S5). Furthermore, learning to predict the EDT is an imitation of human effort on estimating blurry boundary.

The second novelty is that we apply a deep cell detector to generate the markers used in watershed. This deep cell detector determines the final segmented cell numbers with watershed. Introduction of deep cell detector reduces dependency on other techniques like nucleus staining, which makes image acquisition easier. In addition, training on EDT images instead of the original images makes the training much simpler and convergence speed is fast (Fig. S5).

The third novelty of the present work is applying watershed on EDT images for the final segmentation. The deep distance estimator simplifies the original input image and makes it suitable for processing with watershed. Even though the Euclidean distance values of blurry boundaries don't equal with that of the clear boundaries in prediction, they are still small and could be the candidates of separation lines in watershed. For instance, the Euclidean distance of blurry boundaries in region 1(Fig. 3A) are larger than the threshold value of the mask, but watershed could still separate the two cells because watershed segments cells based on the trend of spatial change of the input (Fig. 4D). The training labels of blurry boundaries are probably not precise due to the limit of manual labeling, but the absolute values are less important than the relative values (gray level change). The training accuracy of deep distance estimator is only around 50% (keras metrics: *mae*), but the final CCA could reach 90% because the final segmentation is performed with watershed. Thus, watershed algorithm reduces the accuracy's reliability on training data, which also reduces the influence of inaccurate information in training data.

The aim of single cell segmentation is to detect enclosed boundaries of individual cells. In practice, cell boundaries are often not clear in microscopic images. Several factors contribute to blurred images. First, the image resolution may not be high enough to provide sufficient information. Second, cells are motile, and in live cell imaging it is unavoidable that some parts are out of focus; this problem becomes more severe in higher-magnification images. In addition, cell boundaries also tend to be blurry when cells undergo cell division and some cell fate change such as apoptosis. These factors bring errors to the FCN-based learning and prediction in either the pixel-wise methods or the combined methods. In this study, we used a relatively small set of manually prepared training data, which may limit the maximum accuracy that can be achieved with various learning methods we tested here. Data augmentation methods like cropping and flipping could improve segmentation accuracy (Perez and Wang, 2017; Taylor and Nitschke, 2017). However, data augmentation methods cannot cover the situation of morphological and phenotypic heterogeneity. For example, cells that undergo mitosis shrink into round shape and detach from the substrate. These cells tend to be out of focus and may overlap with other cells, causing problems in identification and segmentation. Expanding the training data helps to cover the heterogeneous population but preparing the data set is another bottleneck. Manual labeling may introduce artifacts and contaminate the quality of the training set. In general, it is easier to prepare the training set with fluorescence images, but this has its own problems for live cell imaging as discussed above. Therefore, one strategy is to use fluorescence staining to generate a data set for training a neural network to segment transmitted light images. A platform for depositing well-curated cell images can benefit the community for training and comparing different algorithms. The problem of limited training set can also be ameliorated with recent development of transfer learning to train CNNs to recognize cells of types different from those in the training set (Shin et al., 2016; Yosinski et al., 2014) and techniques for generating training data such as semi-synthetic method (Weigert et al., 2018).

As discussed in the Introduction, it is desirable to take transmitted light images alone due to photo toxicity concerns and increased effort requirements in some applications. In other cases, one may acquire both transmitted light channel and fluorescence channel images simultaneously. In principle, these two types of images could provide complementary information to facilitate cell segmentation. For example, one may use image composition, with the composite image containing information from both images. In this case, more training data is needed to obtain global optimal parameters of CNN. Future studies may test the combined deep learning and watershed segmentation procedure presented in this work using new CNN architectures with multi-inputs, such as having both transmitted light channel and fluorescence channel images (Raza et al., 2017).

## 5. CONCLUSIONS

In conclusion, we developed a general method for segmenting single cells of images including fluorescence, phase contrast, and DIC. Compared with previous pixel-wise classification methods based on FCN, our method shows higher CCA, which is critical for single cell analysis and cell tracking. It does not require exhaustive parameter tuning and gives prediction on segmentation with high accuracy.

**Abbreviations**

Adam: Adaptive Moment Estimation

CCA: cell count accuracy

CNN: convolutional neural networks

CWCE: class weighted cross entropy

DIC: differential interference contrast

DMEM: Dulbecco's Modified Eagle Medium

EDT: Euclidean distance transform

EGFP: enhanced green fluorescent protein

FBS: Fetal Bovine Serum

FCN: fully convolutional networks

FP: false positive cell count

IU: intersection over union

MSE: mean squared error

RELU: rectified linear unit

ROI: classifier of region of interest

RPN: region proposal network

R-CNN: Region with CNN

TP: true positive

VGG: Visual Geometry Group

# DECLARATIONS

**Ethics approval and consent to participate**

Not applicable

**Consent to publish**

Not applicable

**Availability of data and materials**

The data and program in current study are available in the link

[https://github.com/opnumten/single_cell_segmentation](https://github.com/opnumten/single_cell_segmentation)

**Competing interests**

The authors declare that they have no competing interests.

**Funding**

This work was supported by the National Science Foundation [DMS-1462049], NCI [R01CA232209], and NIDDK (R01DK119232) to JX. We would like to acknowledge the NIH supported microscopy resources in the Center for Biologic Imaging at University of Pittsburgh, specifically the confocal microscope supported by grant number 1S10OD019973-01.

**Authors' Contributions**

WW, JX conceived the project. WW performed the experiment and data analysis. DT contributed in data analysis. YC, JZ, CW and SW contributed in acquisition of data. WW and MX contributed in algorithm design. WW, DT, MX, and JX wrote the manuscript. All the authors read and approved the manuscript.

Meijering, E., Dzyubachyk, O., Smal, I., and van Cappellen, W.A. (2009). Tracking in cell and developmental biology. Seminars in cell & developmental biology *20*, 894-902.

Meijering, E., and van Cappellen, G. (2006). Biological image analysis primer. Erasmus MC, Rotterdam.

Mullassery, D., Horton, C.A., Wood, C.D., and White, M.R. (2008). Single live-cell imaging for systems biology 9. Essays in biochemistry *45*, 121-134.

Muzzey, D., and van Oudenaarden, A. (2009). Quantitative time-lapse fluorescence microscopy in single cells. Annual Review of Cell and Developmental *25*, 301-327.

Noh, H., Hong, S., and Han, B. (2015). Learning deconvolution network for semantic segmentation. Proceedings of the IEEE International Conference on Computer Vision, 1520-1528.

Panchapagesan, S., Sun, M., Khare, A., Matsoukas, S., Mandal, A., Hoffmeister, B., and Vitaladevuni, S. (2016). Multi-task learning and weighted cross-entropy for DNN-based keyword spotting. INTERSPEECH, 760-764.

Perez, L., and Wang, J. (2017). The effectiveness of data augmentation in image classification using deep learning. arXiv preprint arXiv:171204621.

Raza, S.E.A., Cheung, L., Epstein, D., Pelengaris, S., Khan, M., and Rajpoot, N.M. (2017). Mimo-net: A multi-input multi-output convolutional neural network for cell segmentation in fluorescence microscopy images. 2017 IEEE 14th International Symposium on Biomedical Imaging (ISBI 2017), 337-340.

Redmon, J., Divvala, S., Girshick, R., and Farhadi, A. (2016). You only look once: Unified, real-time object detection. Proceedings of the IEEE conference on computer vision and pattern recognition, 779-788.

Ren, S., He, K., Girshick, R., and Sun, J. (2015). Faster R-CNN: Towards real-time object detection with region proposal networks. Advances in neural information processing systems, 91-99.

Roeder, A.H., Cunha, A., Burl, M.C., and Meyerowitz, E.M. (2012). A computational image analysis glossary for biologists. Development *139*, 3071-3080.

Roerdink, J.B.T.M., and Meijster, A. (2000). The watershed transform: definitions, algorithms and parallelization strategies. Fundam Inf *41*, 187-228.

Ronneberger, O., Fischer, P., and Brox, T. (2015). U-net: Convolutional networks for biomedical image segmentation. International Conference on Medical Image Computing and Computer-Assisted Intervention, 234-241.

Shin, H.-C., Roth, H.R., Gao, M., Lu, L., Xu, Z., Nogues, I., Yao, J., Mollura, D., and Summers, R.M. (2016). Deep convolutional neural networks for computer-aided detection: CNN architectures, dataset characteristics and transfer learning. IEEE transactions on medical imaging *35*, 1285-1298.

Simonyan, K., and Zisserman, A. (2014). Very deep convolutional networks for large-scale image recognition. arXiv preprint arXiv:14091556.

Su, H., Yin, Z., Huh, S., and Kanade, T. (2013). Cell segmentation in phase contrast microscopy images via semi-supervised classification over optics-related features. Medical image analysis *17*, 746-765.

Taylor, L., and Nitschke, G. (2017). Improving deep learning using generic data augmentation. arXiv preprint arXiv:170806020.

Uchida, S. (2013). Image processing and recognition for biological images. Development, growth & differentiation *55*, 523-549.

Van Valen, D.A., Kudo, T., Lane, K.M., Macklin, D.N., Quach, N.T., DeFelice, M.M., Maayan, I., Tanouchi, Y., Ashley, E.A., and Covert, M.W. (2016). Deep learning automates the auantitative analysis of individual cells in live-cell imaging experiments. PLoS Comput Biol *12*, e1005177.
29